# Laboratory-Tutorial activities for teaching probability

Michael C. Wittmann, Jeffrey T. Morgan, Roger E. Feeley
Department of Physics and Astronomy, University of Maine, Orono ME 04469-5709
*email*: wittmann@umit.maine.edu; *tel*: (207) 581–1237; *fax*: (207) 581–3410

## Abstract

We report on the development of students' ideas of probability and probability density in a University of Maine laboratory-based general education physics course called *Intuitive Quantum Physics*. Students in the course are generally math phobic with unfavorable expectations about the nature of physics and their ability to do it. We describe a set of activities used to teach concepts of probability and probability density. Rudimentary knowledge of mechanics is needed for one activity, but otherwise the material requires no additional preparation. Extensions of the activities include relating probability density to potential energy graphs for certain "touchstone" examples. Students have difficulties learning the target concepts, such as comparing the ratio of time in a region to total time in all regions. Instead, they often focus on edge effects, pattern match to previously studied situations, reason about necessary but incomplete macroscopic elements of the system, use the gambler's fallacy, and use expectations about ensemble results rather than expectation values to predict future events. We map the development of their thinking to provide examples of problems rather than evidence of a curriculum's success.

# I. INTRODUCTION

Though probability and simple statistics are common in our society (ranging from baseball batting averages to polling data with its statistical measures) we rarely teach the concepts in our introductory physics courses. The little research carried out on teaching probability has been in the context of quantum physics.[1-3] In *Intuitive Quantum Physics*, a general education physics course at the University of Maine, members of the Physics Education Research Laboratory (in alphabetical order: Katrina E. Black, David E. Clark, Roger E. Feeley, Jeffrey T. Morgan, Eleanor C. Sayre, and Michael C. Wittmann) have created a set of activities for teaching probability in a way that can help students later learn concepts of quantum physics. We describe the effect of our course on student understanding of probability and probability density in classical situations.

The goal of our course has been to make physics concepts accessible to students who are on their way to learning the introductory ideas of quantum physics. We use the Copenhagen interpretation of the wave function in *Intuitive Quantum Physics*, and students are required to understand that $|\Psi|^2$ gives the local probability density of finding a particle at a given location. To help them understand this concept, we begin by teaching about probability in non-physics situations and then move into discussions of probability density in physical situations such as a ball toss or oscillating glider.

In this paper, we describe the curriculum so that others may adapt it to their own needs. We describe student reasoning while learning from the curriculum to help inform the choices made by those adapting the materials. The teaching materials are accessible at [4] and we only summarize them in this paper. Research data are given based on the Fall, 2005 semester in which we were best able to track student learning on the topic of probability density. Results are consistent with results gathered from previous semesters in which we do not have as comprehensive a set of longitudinal data. We describe data from several ungraded pre- and post-tests and two examination questions, as well as informal observations which are consistent with our research data.

# II. STUDENT DEMOGRAPHICS AND COURSE DESIGN

The University of Maine, a Carnegie research I institution, is the flagship campus of the Maine state university system. The Department of Physics and Astronomy offers a course in "Descriptive Physics" which is taken primarily by non-science majors needing a laboratory science course for core-curriculum "general education" graduation requirements. The course format was originally designed by R.R. Harrington.[5] The course is taught with three hours of laboratory time and three hours of lecture time. All lab time is placed between the first and second lectures in a given week. Class sizes in this course are typically between 40 and 70. In Fall, 2005, 47 students began the course and 43 completed it.

Students signing up to take the Descriptive Physics course typically have unfavorable expectations and attitudes about physics. Informal results show that students enter the course expecting (and worried about) the mathematical nature of the course. More formally,[6] we find that they enter the course expecting to use "memorize and repeat" learning methods and do not have a view of physics that includes conceptual understanding. Based on several semesters of data using the Maryland Physics Expectations Survey 2 (MPEX 2),[7, 8] we find that students typically enter the course having roughly 50% unfavorable and 25% favorable expectations about conceptual learning. These scores are substantially lower than a typical introductory class, as measured by the original MPEX.[8] The high number of neutral responses is consistent with our observations of other non-science-major courses at the University of Maine or high school courses.

We have developed the *Intuitive Quantum Physics* course to best match student concerns and abilities while teaching them a meaningful physics course on an interesting topic. Starting with optics and waves, students develop simple ideas of quantum physics while continually connecting ideas to their everyday life. We use as little algebra as possible and emphasize other valid methods of reasoning, such as graphical analysis and qualitative reasoning. Where possible, we have students build an understanding of physics from easily observable phenomena, giving students touchstone concepts when dealing with more complicated topics. For example: superposition concepts are always chained back to the observed superposition of waves on a spring; wave and particle interpretations of quantum particles are chained back to the 2-slit interference situation in which individual particles arrive on a screen and eventually fill in an interference pattern; ramped potential energy diagrams are chained back to carts on ramps and square well potential energy diagrams are chained back to a cart bouncing back and forth "frictionlessly" and perfectly elastically between two hard walls. Most material is introduced in a single, 3 hour long lab-tutorial period. The lab-tutorial includes individual, small group, and large group (full class) activities.

The *Intuitive Quantum Physics* course is split into three units in which students develop skills prerequisite to understanding quantum physics, create a "toolbox" with which to study the quantum world, and discuss applications of quantum physics to the real world. Table I summarizes the course structure. We provide this information to situate the activities on probability within the larger context of the course. We teach probability to provide a language which lets students understand the behavior of particles in the wave-particle duality two slit experiment: where might the next particle arrive on screen? The ideas taught are used to discuss probability density in bound state problems and other quantum physics situations.

|  |  | **Nature of Science questions** | | |
|---|---|---|---|---|
|  |  | **How do you know?** | **How can you explain?** | **Why do you believe?** |
| **Content Unit** | **Unit 1: Optics and Wave Physics**<br>• Light travels outward in a straight line from every source point<br>• Superposition of visible waves on a spring and in water<br>• Water and light interference<br>• Wave-particle duality – the great dilemma! | "I saw it" | "It's like something in real life" | "I saw it" |
|  | **Unit 2: A New Toolbox**<br>• Energy diagrams<br>• Classical probability<br>• Curviness of graphical functions<br>• A graphical interpretation of the Schrödinger equation | "You told me" | "It's like something else I know" | "I thought about it" |
|  | **Unit 3: Topics in Quantum Physics**<br>• Quantization in finite wells; bound states<br>• Spectroscopy<br>• Models of molecules<br>• Quantum tunneling | "I figured it out" | "It's consistent with these other things" | "I'm not sure I do, but I can think about it, anyway" |

TABLE I: Course outline. Each unit has specific goals concerning content knowledge and the nature of science.

## III. TEACHING PROBABILITY

To summarize the instructional materials for teaching probability: we introduce concepts of probability by first looking at discrete, non-physics systems (such as coins, and dice) and then study macroscopic physical systems (such as balls tossed in the air or oscillating air gliders). Later, students connect these ideas to potential energy graphs for the same physical situations – they must learn to connect potential and kinetic energy and consider where the object is moving slowest and spends the most time.

As discussed in the next section, we find that students enter our courses unable to apply many of the ideas about probability. Many apply the gambler's fallacy to situations, while others are unclear about the interpretation of probability density and the link between an object's speed and the likelihood of finding it in a region of space.

The primary ideas we wish students to learn are:
- the sum of probabilities of outcomes to an event equals 1 (or 100%),
- expected distribution results can be predicted, but most likely do not match actual outcomes of ensembles of events,
- in physical systems, we can compare the time in one region of space to the total time in all regions of space as a way of finding the probability of being located in that region,
- relative time in a region of space is determined by the speed of the object in that region.

We visit these concepts twice, in lab-tutorial 5 (Probability) and 7 (Probability and Energy). In lab-tutorial 6, we introduce concepts of kinetic and potential energy, create potential energy diagrams for several "touchstone" systems (harmonic oscillators, ramps, square wells, and barriers) and do not discuss probability or probability density. The descriptions below come primarily from lab-tutorial 5 (Probability) unless noted.

### A. Discrete systems

We introduce the idea of probability by asking students to consider a person reaching into a box containing 10 balls (five are checkered, three are striped, and two are solid) and removing a ball. Students are asked to find the probability of picking each type of ball and finding the total probability. We thus establish the idea of a probability of 1 (or 100% chance of an event occurring, in this case "a ball being picked"). Students typically have no problem with this idea.

Students follow up with a series of questions about coin tosses. After 10 tosses, students expect 5 heads and 5 tails. Is 6 of one and 4 of the other a surprise? Is 10 of one and 0 of the other a surprise? Students carry out 10 coin tosses individually and then work in groups to compare answers. A range of results helps them discuss variability of results.

Students are next asked to toss three coins at once. Results will be variations of HHH (3 heads, no tails), HHT, HTT, and TTT. Coin order matters, and students are asked to predict all possible outcomes and discuss the probability of finding each. At this point, we introduce a histogram representation, shown in Figure 1.

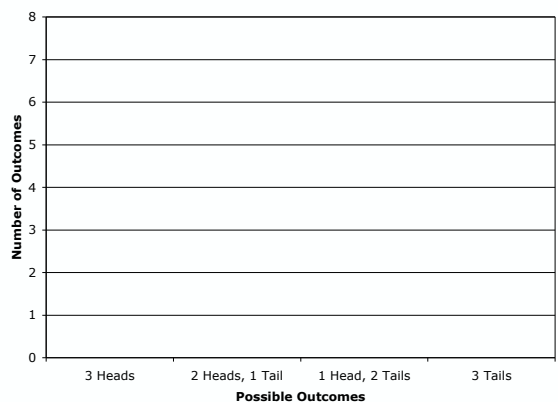

FIGURE 1: Probability histogram chart for the 3 coin toss.

Once students have predicted results, they are asked to carry out the experiment and toss 3 coins in unison a total of 8 times. Students compare a histogram of their actual data to the prediction. Results vary from student to student. Again, there is an opportunity to discuss the difference between predictions and results. Students also discuss the role of larger ensembles of results after combining all results from a table.

Students move on to consider distinguishable dice (2 differently sized or colored dice are used in class). They roll the dice together, keeping track of each individual die's result and the sum of the two. Again, a probability histogram is created and its results are compared to the 36 possible combinations of dice tosses.

### B. Physical systems

Students move on to a discussion of physical systems, including a ball tossed into the air and a glider oscillating on an airtrack when attached by springs to supports at the edge of the airtrack.

For the ball toss, students must predict if a ball thrown in the air is more likely to be observed at the top, middle, or bottom third of its trajectory (labeled A, B, and C, see Figure 2). Students must interpret video of physical situations, counting the number of frames an object spends in a given region of space. In the process, we have them extend a previous representation[1, 9] in which strobe photographs were considered. They bridge to the idea of probability density as the proportion of the total time spent in a region of space by comparing the number of video frames spent in a region to the total number of frames. Then, to ask for the probability of finding an object in one region of space over another, one need only consider the ratio and compare. Students qualitatively compare the speed of the ball to the probability of finding it in a region of space and find that it is most likely to be where it is moving slowest. Students often expect it to be at the bottom ("where it lands") or in the middle ("because it goes through it twice"), so the result is surprising to many.

Students end the lab-tutorial with a discussion of a harmonic oscillator. Using an air glider which is attached by springs to supports,

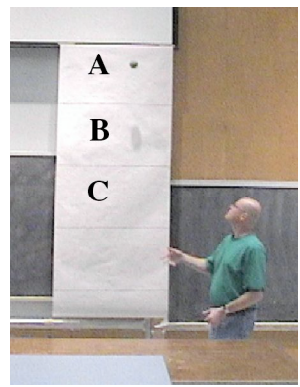

FIGURE 2: Ball toss problem for discussing the probability of finding a ball in a given region when thrown vertically

we have created a very low friction situation in which students can study probability. Again, we "bin" the physical system into regions (see Figure 3). Students use results from the ball toss problem to predict the probability of being found in a given location. Students are asked to respond to a fictitious student dialogue in which common misunderstandings about probability[10, 11] are used as some elements of reasoning while correct ideas are scattered throughout the dialogue. Incorrect ideas include the statement that region C is passed through more often and is therefore more likely, that region C is where the cart "wants to end up" and is therefore more likely, and the idea that each region is equally likely because it is equally wide.

Once students have predicted what they will observe, they count frames in a videotape of the situation (video available at http://perlnet.umaine.edu/abt/v2video.htm). They use a probability histogram to plot their results. They also must compare their results to their interpretation of the student dialogue.

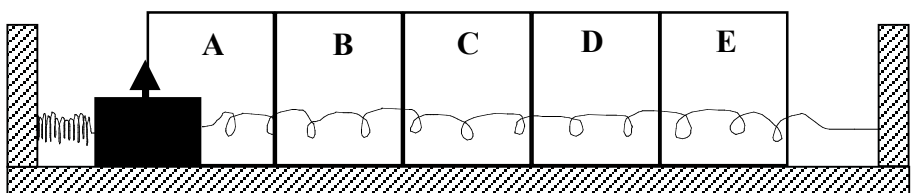

FIGURE 3: Harmonic oscillator system. An air glider is attached by springs. The region of oscillation is split into 5 bins.

In lab-tutorial 7 (Probability and Energy), students revisit the ideas from lab-tutorial 5 (Probability) but interpret them more formally. First, they have developed the idea of the harmonic oscillator potential energy graph, $PE = 1/2\, kx^2$. Second, they define the concept of probability density more formally. Students are given the histogram which they found in lab-tutorial 5 and asked to discuss the relationship between potential energy, kinetic energy, and probability density.

Because the concepts of probability and probability density are revisited, we have a rich environment for studying the development of their learning. Pre-tests for lab-tutorial 7 serve as post-tests for lab-tutorial 5, for example. Data from investigations into student reasoning are given below. They are meant to illustrate the issues when teaching probability and not to make claims about effectiveness of instruction or consistency of responses.

### C. Extending probability histograms to new situations

At the end of the lab-tutorial 7 (Probability and Energy), students participate in a full-class discussion in which they are asked to compare probability histograms from the coin toss, dice rolls, ball toss, and oscillating cart to other situations. These include the probability histogram for a frictionless cart bouncing between two perfectly elastic walls (i.e. no energy is lost in the system) and the probability histogram for electrons in a low intensity 2 slit experiment (found in a previous week's activity, see Figure 4). This same histogram was used at the beginning of lab-tutorial 5 (Probability) to motivate a discussion of probability in the week after students first studied wave-particle duality.

## IV. STUDENT UNDERSTANDING OF PROBABILITY

Having described the teaching sequence used to help students develop an idea of probability density, we now describe students' ideas as they enter the course and how these ideas develop during instruction. We use evidence from Fall, 2005, since students in that semester were most extensively studied.

### A. Understanding ensembles of easily understood events

Before instruction on probability, we asked students a pre-test in which all knew that the likelihood of a coin toss was 50-50, heads or tails. We then asked students a multiple choice question (shown in figure 5) in which we asked students to explain their reasoning.

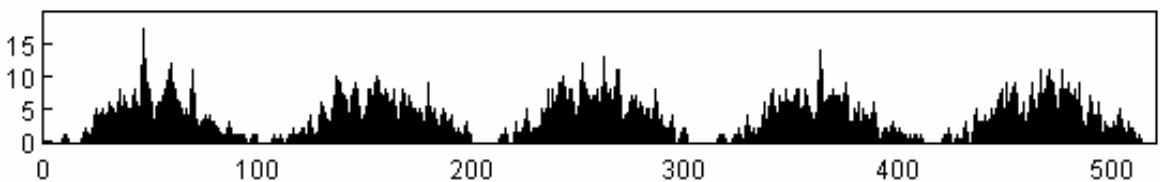

FIGURE 4: Histogram of electron frequency on a screen after a simulated 2-slit interference experiment. Note similarity to histogram representation in Figure 1.

---

A penny has been flipped 10 times, with the following results.
Heads, Tails, Tails, Heads, Heads, Tails, Heads, Heads, Heads, Heads
On the 11th flip, the likelihood that the penny will be "heads" is…
a.　　…much greater than 50-50.
b.　　…slightly greater than 50-50.
c.　　…50-50.
d.　　…slightly less than 50-50.
e.　　…much less than 50-50
Explain your reasoning.

FIGURE 5: Probability question. The question was designed to elicit the gambler's fallacy and other issues related to expectation values.

Our reason for asking the question with 10 previous tosses was to help students work out the ratio of 70% heads and 30% tails. We used a 70-30 ratio because 60-40 is too close to 50-50 and perhaps not sufficiently different from the expected 50-50. We wished the result of the first 10 tosses to be unexpected but not outrageously so (as ten consecutive tosses of Heads would have been). A correct answer is (c), 50-50, with the reasoning that previous tosses of the coin have no effect on an event. Thus, for a fair coin, one can both ignore the ensemble of past results and also not compare the ensemble to the eventual, expected result. A majority of students gave the correct answer, but many gave other responses (see Table II). The results show that many students enter our courses with everyday expectations about what the ensemble values of sets of coin tosses should be and use past results to decide future events. Our results are consistent with those found by Bao when discussing similar problems.[12]

B. Interpretations of probability density for physical systems

We asked a series of questions (often identical) throughout the semester to investigate the development of student reasoning about probability density in physics systems. We note that pretests and post-tests were never handed back to students, and only the tutorials and tutorial homework were in students' hands as they studied for examinations.

*1. In a gravitational setting*

Before any instruction on probability, students were asked the question shown in Figure 6 on an ungraded lab-tutorial pretest. The goal of the question was to find what methods students used for determining the likelihood of finding an object in space. Students are told that the falling raindrop is constantly speeding up, so it is most likely to be in region A where it spends the most time.

We discuss three answers from this pretest. One-third of the 42 students answering the question gave the right answer with the correct reasoning. We group other responses into two major categories.

*Random observation of equal sized bins means equal likelihood.* Slightly more than 1/3 (15 of 42) of the students said that the drop was equally likely to be found in each region. The reasoning given was either that equal sized bins lead to equal probabilities or that the random nature of the observation made each region equally likely. Neither explanation was found again after instruction, though the former may have been helpful when discussing the "free cart," described below.

*Starting or ending issues.* Roughly 1/5 (8 of 42) of the students described the most likely scenario as either region A or C because the drop was starting or ending there, respectively. We believe that these students are not looking at the entire motion of the drop and therefore cannot make the comparison of "time in a region" to "total time for drop to fall." Thus, they use other reasoning, in this case the idea that starting or ending points are more likely to be observed. We note that another 2 students (5%) described B as the most likely area because it is in the middle. This response was given more commonly in classroom situations in lab-tutorial 5, when a ball tossed up and down was considered; students described the ball passing through region B twice, for example. Results are similar to those by Ambrose with more advanced students.[10, 11]

| 57% | 24% | 14% | 7% |
|---|---|---|---|
| Expect 50-50 no matter what came before. | Getting too many of a kind in a row is unlikely: 4 heads in a row seems unlikely, so tails must come soon. | Results should end up at 50-50, so tails is likely in the future | Bayesian interpretation: since past results are 70-30 H-T, we should continue to have that ratio in the future. |

TABLE II: Student responses and typical explanations on the 11[th] coin toss problem. N = 42. One student gave two different explanations, so totals do not add to 100%.

An extremely light rain is falling out of a SmartCloud, such that only one drop of rain is in the air at a time, and as soon as one drop hits the ground, another is released from the cloud. (See the picture below.) Each drop speeds up continually until it hits the ground. The space between the cloud and the ground is divided into three equal-size regions, A, B, and C. 100 pictures are taken of this system at random times. The majority of the photographs will show the raindrop in…
a. Region A
b. Region B
c. Region C
d. There will be about an equal number of each.
Explain your reasoning

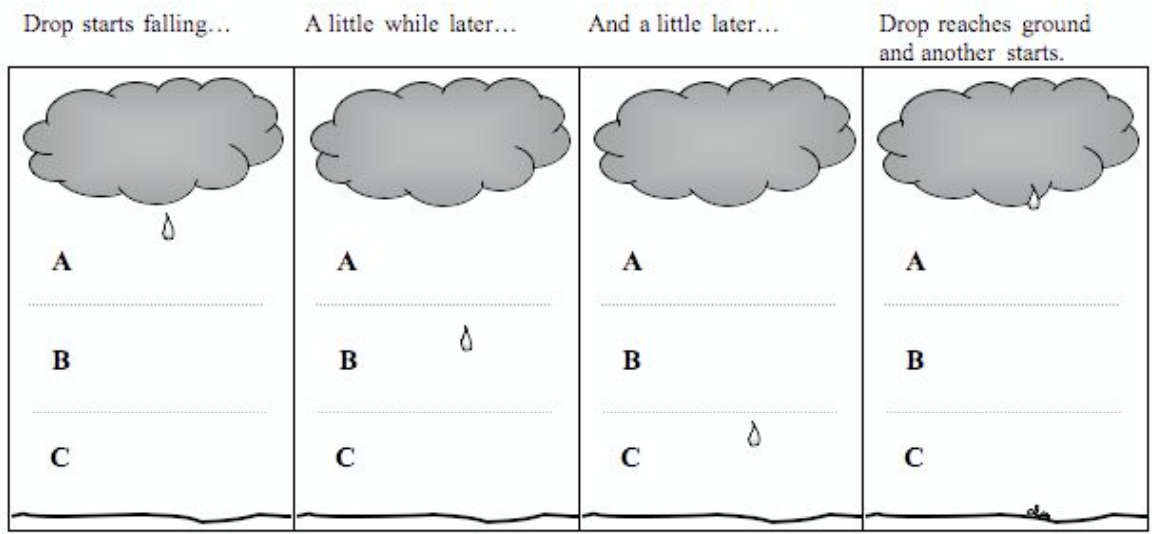

FIGURE 6: Pretest question on probability density. Students were asked to predict where a single falling raindrop (constantly speeding up) was most likely to be observed.

*2. The "free cart" between hard walls*

We refer to a cart bouncing back and forth between two hard walls and never losing energy as the "free cart," in comparison to a cart attached to springs (the "harmonic cart") or a cart on a ramp. Students studied the harmonic cart in lab-tutorial, and we chose to ask questions about a simpler scenario when pre- and post-testing them. The basic question on the three tests was essentially identical (see Figure 7), though the post-test and examination contained additional elements which are mentioned below, as needed.

The pre-test was given the week before lab-tutorial 7 (Probability and Energy). As noted above, it served in part as a post-test of student learning in lab-tutorial 5 (Probability). The post-test was given as an ungraded quiz on the first day of lecture after all students had completed lab-tutorial 7. The examination was given six weeks later, near the end of the semester. During those weeks, students had worked primarily on the interpretation of probability density in the context of the Schrödinger equation. There was lecture discussion on the differences in probability density between an electron in the ground state of a finite square well and the free cart between hard walls, so the topic was revisited in lecture after lab-tutorial instruction.

Data from the three tests are shown in Table III. Note that we include only the two most common responses and do not include the rare other responses in the case of the ungraded post-test. Students giving the correct response (equally likely in all regions of space) also gave correct reasoning by saying that the speed is the same in all regions and the time spent in all regions is equal. We describe four specific results from this table.

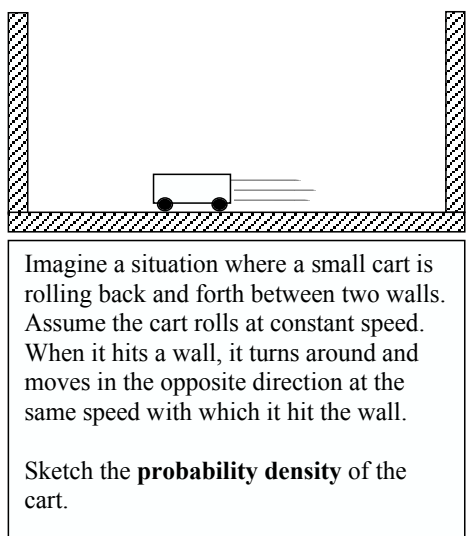

Imagine a situation where a small cart is rolling back and forth between two walls. Assume the cart rolls at constant speed. When it hits a wall, it turns around and moves in the opposite direction at the same speed with which it hit the wall.

Sketch the **probability density** of the cart.

FIGURE 7: "Free cart" probability density question.

|  | Pre-test N = 38 | Post-test N = 41 | Examination N = 43 |
|---|---|---|---|
| Equal in all | 52 % | 52 % | 88 % |
| higher at edges | 33 | 48 | 10 |

TABLE III: Student responses on the free cart problem.

*Hidden variables in the prevalence of correct answers*. We do not have sufficient data to understand either the lack of change from pre- to post-test nor the increase in results from the post-test to the examination and believe several variables may play a role. The post-test was administered the day after laboratory; students had not yet completed the homework and had not had time to practice the ideas. On the homework, students had to sketch the probability density for a runner moving at different speeds. Furthermore, the tutorials on tunneling included a discussion of the "magnetic cart," in which a cart with magnets attached passes through a potential barrier made by a strong magnetic field. In addition, instruction during lecture time emphasized comparisons between the quantum particle in a finite square well and the "free cart." Finally, it is possible that the difference in students attitudes toward an ungraded post-test and a graded examination question also played a role. Each of these may have had an effect in increasing student understanding of probability density.

Three different issues arise when analyzing the increase in the incorrect response in the post-test.

*Pattern matching*. From the student responses, it seems that many on the pre- and post-test were pattern matching to the harmonic oscillator which had already been discussed in lab-tutorial 5 and was again discussed in lab-tutorial 7. We note that our analysis does not include those students who correctly discussed carts slowing down to turn around; their graphs were notably different from those who pattern matched to the harmonic oscillator. The responses we counted as "higher at edges" were mostly parabolas or similar shapes, some actually identical to the histograms drawn in lab-tutorial. We believe students are pattern matching to previously taught material without considering the actual situation. That more pattern match after lab-tutorial 7 than before is worrisome, but consistent with the expectations that many students have [6] that one learns physics by memorizing and not by thinking through a situation.

*Counting issues and edge effects*. Many students in a classroom discussion following the post-test supported the idea that the free cart was more likely to be found at the edges because it was in that region of space both on its way to the wall and again after bouncing off. It was there twice for every one time in another region. Other students during the class discussion made the correct point that one needed to compare "time in a region" to the accurate "total time," namely for a full cycle of the cart back and forth in *both* directions. Thus, the cart was in each region twice, not just those regions at the edges. That the cart was in regions at the edges twice in a row did not matter. This point was accepted in the large-lecture discussion, but may have been caused by activities in the lab-tutorial itself. When counting frames in a video, one counts all frames for a region at the edge consecutively, and might treat the regions away from the edges differently.

*Limited types of responses*. Finally, we note that no other responses than those shown in the table "survived" the instructional process. Though the pre-tests contained some other (not shown) responses, these were no longer given after instruction. The lasting issue for students remains one of understanding the physical situation. Either students pattern-matched to the wrong system or they had problems understanding which region of space to consider. It is also possible that students switched from one wrong answer to another. We note that 8 of the 22 students who gave the correct answer on the pre-test switched to the incorrect answer of "higher at the edges" on the post-test. With two, it is clear they were pattern matching to the harmonic oscillator situation, while the sketches and explanations of the other six do not allow us to distinguish what guided their reasoning.

### 3. The ball on stepped ramps

On the final examination, students answered a question taken from Jolly and Bao [1,9] (see Figure 8). In the question, students must analyze the speed of the object on different levels, recognize that the levels are of equal length, and compare the time on each level to the total time for a ball to traverse the path. A correct answer would show $P(x)$ twice as high from 0 to L as from L to 2L.

We counted qualitatively correct responses (without the explicit 2-to-1 ratio) as correct in our analysis, as long as the students' explanations qualitatively correctly described relative speeds and times on each level, but list them separately in Table IV. We did not accept student responses in which they seemed to map the physical picture to the ramps to the $P(x)$ graph. Of the 44 students answering the question, 83% were completely or qualitatively correct with correct reasoning, while 14% showed some sort of sloped line, usually on level 2. Other responses (not listed in the table) were given by only a one or two students at a time.

|  | Final exam |
|---|---|
| Correct with 2:1 ratio | 45 % |
| Qualitatively correct | 38 |
| sloped line in one region | 14 |

TABLE IV: Student responses on the final examination balls-on-ramps question, N = 44.

Consider the experiment shown below. A series of balls is set moving towards the right at a **very small velocity** $v_0$. (Ignore friction.) On level 2, a ball moves twice as fast as on level 1.

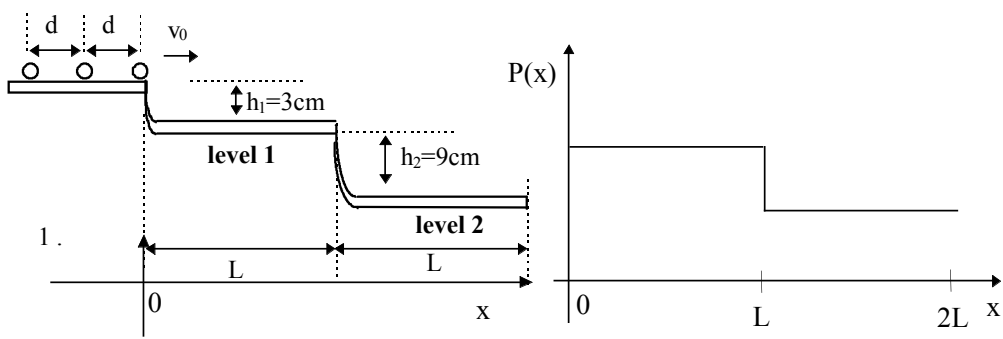

1. On the axes above, sketch the **probability density** of finding a ball.

2. Explain how you arrived at the shape that you have drawn in question 1.

FIGURE 8: "Ball on ramps" probability density question. Students answered this question on the final examination. A correct graph is shown on the right.

## V. DISCUSSION OF RESEARCH AND INSTRUCTION

When describing the likelihood of finding a particle in one region of space over another, students must reason about time in a region compared to the total time for an event to occur, meaning they must understand the physical system well enough to compare speeds in regions of space and to understand how long the event takes. Our teaching materials are designed to help students develop these ideas. But, students often use other methods to guide their reasoning, even after instruction. They may pattern match to different systems that have been studied in detail. They may focus on consecutive events, such as carts turning around when bouncing off walls. They may also look at specific times during the event, such as its beginning or its end. Finally, students may use correct reasoning but not be able to accurately calculate the times involved. In each case, we can see what they fail to do, but we emphasize what they actually do as a way of better understanding what valuable ideas they have that might be built on in future modifications to our instruction.

We find that there is improvement in student reasoning about probability density, but that teaching only about the classical systems (including graphs of their potential energies) was not sufficient to help students learn the material. Post-test results may have too quickly followed instruction, and show that students did not immediately change their ideas about probability density in a very simple system, the free cart. Later results, on an examination near the end of the semester and a final examination, show that students developed a more complete understanding and that more than 80% were able to answer questions about probability density after additional instruction in simple systems in quantum physics. We note that 29 of the 37 students who correctly answered the examination question about the free cart also answered the balls on ramps question correctly (meaning that nearly 80% were consistent in their correct answers). We believe that having 80% of the class, on a given question, discuss a difficult concept accurately, and to have 80% of that population do so consistently, is an indicator of a successful class, especially when only 1/3 of the students could accurately describe the situation before instruction.

## ACKNOWLEDGMENTS

Katrina E. Black, David E. Clark, and Eleanor C. Sayre contributed to the development of the instructional materials used in the lab-tutorials. We used DataStudio software, from Pasco, but the experiments can easily be created using other, similar software packages. We thank Rachel E. Scherr, John R. Thompson, and R. Padraic Springuel for discussions about and contributions to the course and this paper. Development work on this course is supported in part by NSF grant DUE-0410895.

*Tutorial #5*: **Probability**

*Intuitive Quantum Physics*                                    Name:_____________________________

---

In our last tutorial, we drew and later observed histograms corresponding to the number of electron hits in certain regions of the screen.

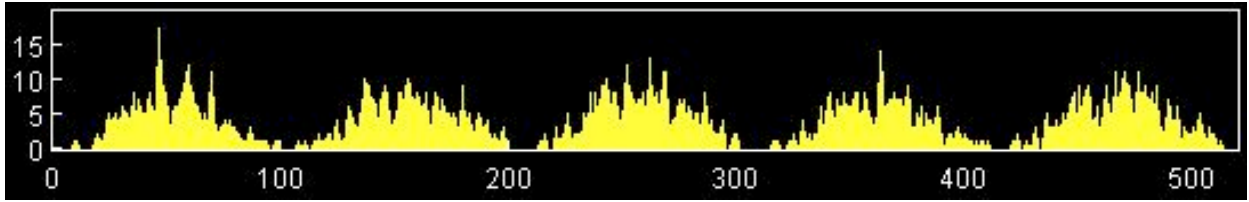

The height of the histogram is related to the probability of an electron landing in a certain location. We need to think a bit more about probability and how it relates to physical systems, which is the point of today's tutorial.

**I.    Introduction to Probability**

We all have some idea of what is meant by the term probability or chance. A student who "guesses" on a true/false problem has a 50-50 chance of getting it right. A TV meteorologist claims there is a 40% chance of rain. A baseball player who has a .279 batting average has a 27.9% chance of getting a hit.

**Probability** is the likelihood that a given event will occur. Imagine a student randomly filling in circles on an answer sheet without even looking at the exam. If each question has four possible responses, the probability that she will choose the correct answer is 1 in 4. We can represent this in several equivalent ways – as a fraction, a number(between 0 and 1), or a percentage(between 0% and 100%):

$$\text{Probability of randomly choosing the correct answer}: \frac{1}{4} \text{ OR } 0.25 \text{ OR } 25\%$$

A.  A box contains 10 balls; five are checkered, three are striped, and two are solid. A person reaches into the box without looking and draws a ball out.

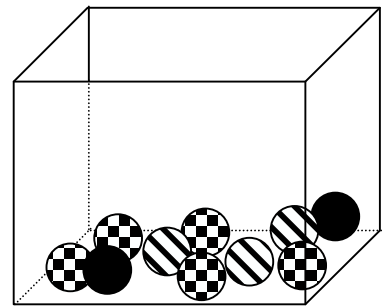

1.  What is the probability that the chosen ball is checkered? Express your answer as a fraction, a number, and a percentage.

2.  What is the probability that the chosen ball is solid? Express your answer as a fraction, a number, and a percentage.

    3. What is the probability that the chosen ball is striped? Express your answer as a fraction, a number, and a percentage.

B. Add the probabilities of choosing a checkered ball, a solid ball, and a striped ball.

## II. Single Coin Experiments

A. Imagine flipping a single penny.

    1. What are the possible outcomes?

    2. What is the probability of each of the possible outcomes?

    3. Add the probabilities of all possible outcomes.

B. Compare your sum of probabilities above to your answer in Part I, B.

    1. In any experiment, what must the sum of all probabilities be?

    2. What is the significance of this number?

C. Now imagine flipping a single penny 10 times. Predict how many heads and tails you think you would observe.

    1. Would you be surprised if the outcome was 6 heads, 4 tails?

    2. Would you be surprised if the outcome was 10 heads, 0 tails?

# Probability

D. <u>Do this section of the tutorial individually</u>. Obtain a penny, and flip it 10 times, recording in the chart below whether the result was *heads* or *tails*.

| Flip: | 1 | 2 | 3 | 4 | 5 | 6 | 7 | 8 | 9 | 10 |
|---|---|---|---|---|---|---|---|---|---|---|
| Outcome: | | | | | | | | | | |

1. How many times was the result *heads*? *Tails*? Express your answers as a fraction of the total number of experiments.

2. How does your data compare to the probability of getting a *head* or *tail*? Are your answers consistent with the probabilities determined above? Were you surprised at the outcome?

E. When everyone has completed Section D, use the entire group's results to answer the following questions.

1. Compare your data with other members of your group. Did everyone get the same results?

2. Add the total number of heads flipped by your group. Express your result as a fraction of the total number of times a coin was flipped.

3. Compare this result with the probability of flipping heads. Discuss any discrepancies.

Though any one group member's data may be closer to the probability of flipping *heads* than the combined result of the entire group, in general the more experiments that are performed, the closer the measured outcome is to the outcome predicted by probability theory.

### III. Multiple Coin Experiments

A. Imagine three pennies are flipped at the same time. What are all the possible outcomes of that event? Use the chart below to organize the outcomes.

| Outcome | Coin 1 | Coin 2 | Coin 3 |
|---|---|---|---|
| 1 | | | |
| 2 | | | |
| 3 | | | |
| 4 | | | |
| 5 | | | |

| Outcome | Coin 1 | Coin 2 | Coin 3 |
|---|---|---|---|
| 6 | | | |
| 7 | | | |
| 8 | | | |
| 9 | | | |
| 10 | | | |

1. How many of the possible outcomes are… (What percentage of the total number of outcomes is this?)

   a. three heads?

   b. two heads and one tail?

   c. one head and two tails?

   d. three tails?

2. Sum all of the percentages. Check that this sum is consistent with the sum of all possibilities for the other scenarios you've examined.

3. Imagine that you flipped 3 coins 8 different times. Use your chart of possible outcomes to construct a histogram representing the number of events as a function of the possible outcomes.

**Theoretical Histogram**

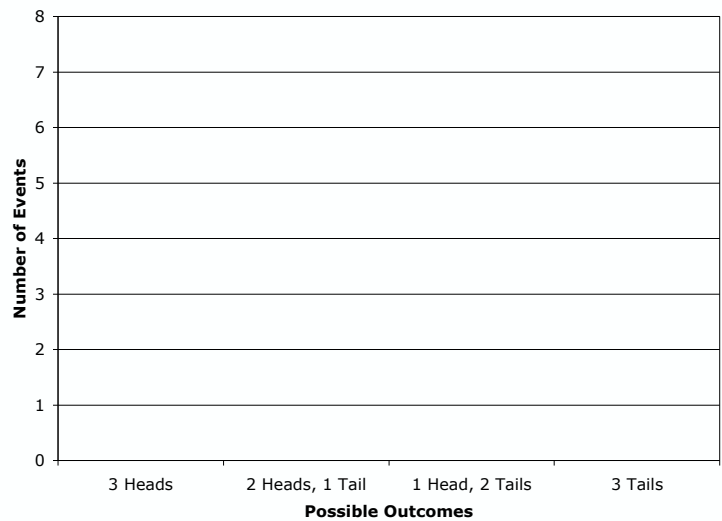

# Probability

B. Obtain 3 pennies. Working individually, flip all three 8 different times. Record the data from your experiment in the table below.

| Event | Outcome |
|---|---|
| 1 | |
| 2 | |
| 3 | |
| 4 | |

| Event | Outcome |
|---|---|
| 5 | |
| 6 | |
| 7 | |
| 8 | |

1. Use your data to construct a histogram representing the number of events as a function of the possible outcomes.

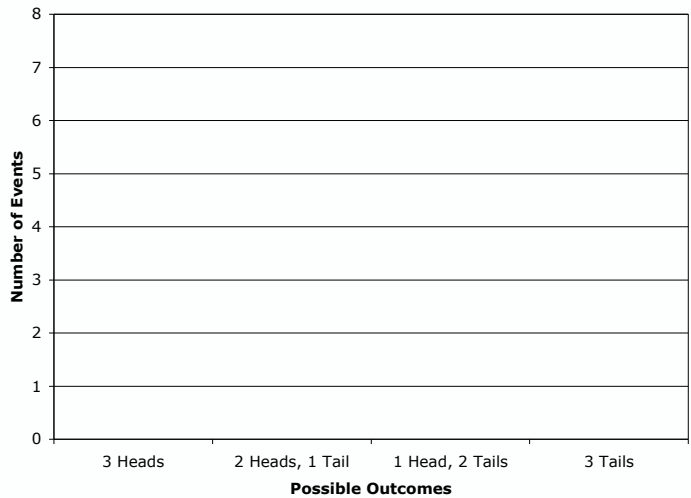

**Experimental Histogram**

2. How does the histogram from your experiment compare with the histograms of other members of your group? Discuss any discrepancies.

3. How does the histogram from your experiment compare with the histogram of theoretical outcomes?

4. What is the difference between an *outcome* and an *event*? Discuss your reasoning with other members of your group.

### IV. Dice Experiments

A. Obtain two distinguishable dice. Record a description of each die to help you track them.

    Die #1: \_\_\_\_\_\_\_\_\_\_\_\_\_\_\_\_\_\_\_\_\_\_\_\_\_      Die #2: \_\_\_\_\_\_\_\_\_\_\_\_\_\_\_\_\_\_\_\_\_\_\_\_\_

Working individually, roll them 20 times, recording the results of your experiment in the table below.

| Roll | Die #1 | Die #2 | Total |
|---|---|---|---|
| 1 | | | |
| 2 | | | |
| 3 | | | |
| 4 | | | |
| 5 | | | |
| 6 | | | |
| 7 | | | |
| 8 | | | |
| 9 | | | |
| 10 | | | |

| Roll | Die #1 | Die #2 | Total |
|---|---|---|---|
| 11 | | | |
| 12 | | | |
| 13 | | | |
| 14 | | | |
| 15 | | | |
| 16 | | | |
| 17 | | | |
| 18 | | | |
| 19 | | | |
| 20 | | | |

B. Draw a histogram for the number of times each total was rolled.

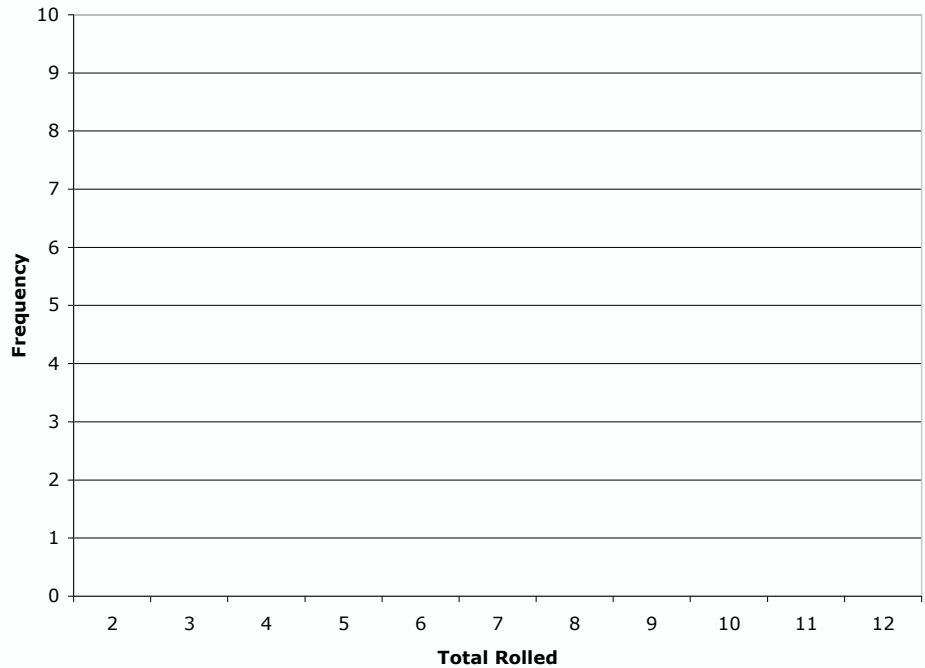

# Probability

C. There are 36 combinations that can be rolled with two dice. The chart below shows the possibilities. They gray boxes represent the possibilities for a single die, the white boxes represent the total of two dice.

|  | | Die 2 | | | | | |
|---|---|---|---|---|---|---|---|
|  | | 1 | 2 | 3 | 4 | 5 | 6 |
| Die 1 | 1 | 2 | 3 | 4 | 5 | 6 | 7 |
|  | 2 | 3 | 4 | 5 | 6 | 7 | 8 |
|  | 3 | 4 | 5 | 6 | 7 | 8 | 9 |
|  | 4 | 5 | 6 | 7 | 8 | 9 | 10 |
|  | 5 | 6 | 7 | 8 | 9 | 10 | 11 |
|  | 6 | 7 | 8 | 9 | 10 | 11 | 12 |

1. How many ways can you roll a total of…

   a.   …2?

   b.   …5?

   c.   …10?

2. Draw a histogram of the probability of rolling each combination.

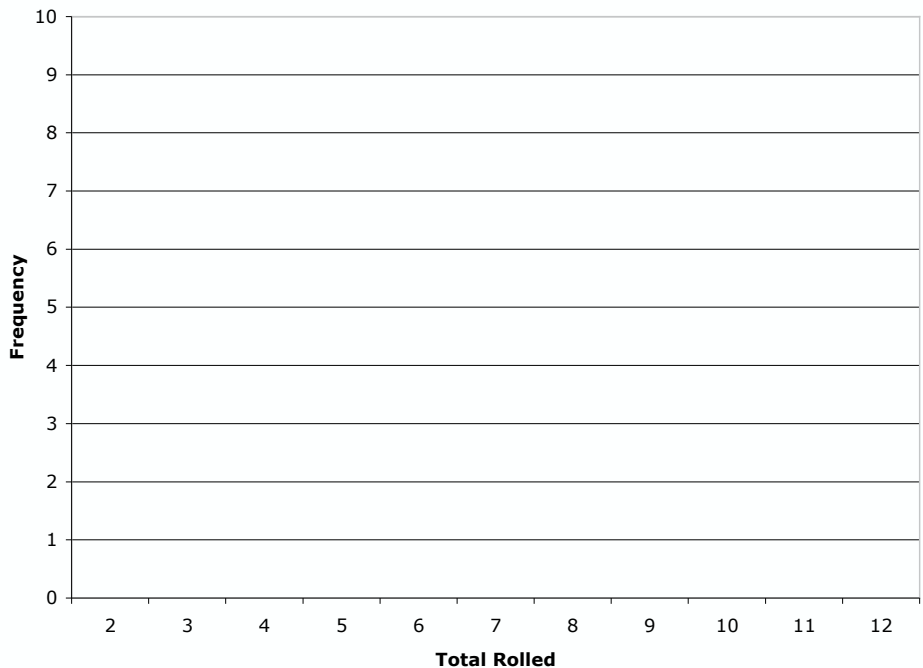

3. Compare the histogram of your data to the histogram of possible outcomes. How are they similar? How are they different?

# Probability

D. Use the chart below to record the number of times each member of your group rolled each total. Add the results for your entire group.

| Total | Group Member 1 | Group Member 2 | Group Member 3 | Group Member 4 | Group Total |
|---|---|---|---|---|---|
| 2 | | | | | |
| 3 | | | | | |
| 4 | | | | | |
| 5 | | | | | |
| 6 | | | | | |
| 7 | | | | | |
| 8 | | | | | |
| 9 | | | | | |
| 10 | | | | | |
| 11 | | | | | |
| 12 | | | | | |

E. Draw a histogram of the number of times each total was rolled by your entire group.

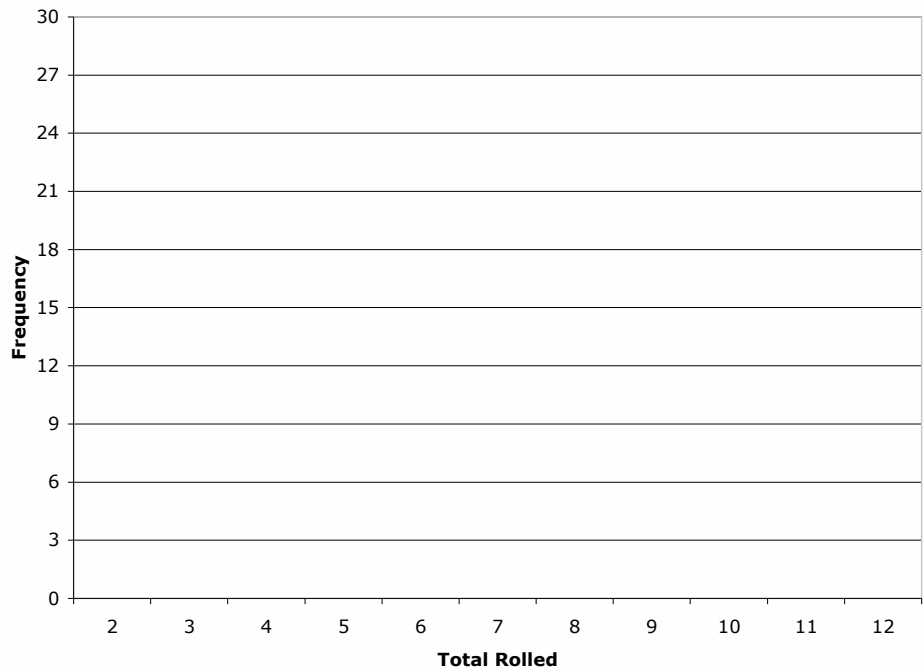

# Probability

**V. Using Probability to Predict**

A. How does the histogram of your group's rolls (page 8) compare to the theoretical histogram of outcomes (page 7)? Is your group's histogram more like the theoretical than the histogram of your rolls alone? Discuss with your group.

B. Imagine tossing your pair of dice one more time. What would be the most likely total rolled? Explicitly state whether your prediction is based on the theoretical model or your data.

C. Imagine tossing your pair of dice twenty more times (in essence, repeating the experiment). Don't do it! Imagine!

   1. Would you expect your results to be identical to your results recorded on page 6?

   2. What would you need to do to check your prediction?

   3. There are several (correct and valuable) interpretations of the word "identical". Decide what the word means to you in the context of question 1, then share your interpretation with other group members.

# Probability TUT5-56

## VI. Board Meeting #1

You (should) know the drill by now.

1.  (Review the end of tutorial 4 if necessary.) Below is the histogram that accompanies the electron experiment in our last tutorial. One electron is shot at the screen. Where will it land?

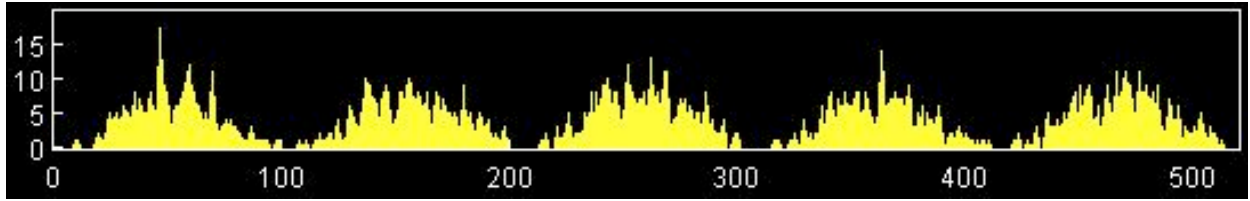

2.  Does the result of a single roll influence the results of other rolls? Explain your reasoning.

3.  Would it be better to predict the results of future rolls based on (a) your table of individual results from page 6, (b) the table of your group's data on page 8, or (c) the chart of theoretical outcomes on page 7? Explain your reasoning.

Discuss the questions in class. Arrive at a consensus on each, and then move on to the next part of the tutorial.

# Probability 

**VII. Tossing a Ball**

A. Imagine a man tossing a ball into the air. The path of the ball is divided into three regions of equal size: A, B, and C. He's perfected his technique, so that the ball leaves the hat at the bottom of Region C, and reaches the peak of its trajectory at the top of Region A.

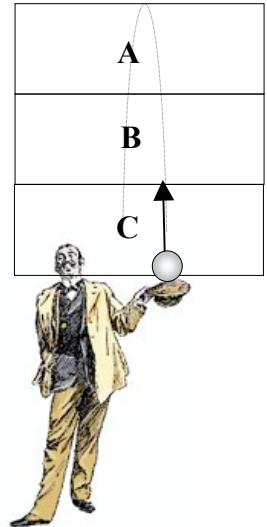

1. Describe the speed of the ball over the course of its motion.

2. If a photograph were taken at a random time, in which region would you expect to find the ball? If all three regions are equally likely, state so explicitly. Explain your reasoning.

3. Lets use some numbers to get a bit more detailed about what is going on.
   If a photograph was taken every 0.01 seconds for 1 second (a total of 100 photographs),

   a. estimate the number of photographs that would show the ball in Region A, in Region B, and in Region C, and

   b. explain your reasoning.

B. The 100 photographs described in part A, question 3 above are placed in a box. Two students predict the outcome of choosing a photograph at random from the box.

   *Student 1: "It's more likely to be a picture of the ball in Region A. The ball spends more time in Region A than it does in B or C."*

   *Student 2: "No, it's equally likely to be any of them. The ball travels the same distance in each region, so there'll be about the same number of pictures from each of the three regions."*

   1. <u>Answer this question individually</u>. With which, if any, of the statements do you agree? Explain your reasoning.

   2. Now discuss both students' reasoning with the members of your group. Do you all initially agree? If not, can you come to a consensus?

# Probability

C. Open the video clip *BallToss.mov* on your computer. Play the movie, and observe Roger tossing a ball in the air.

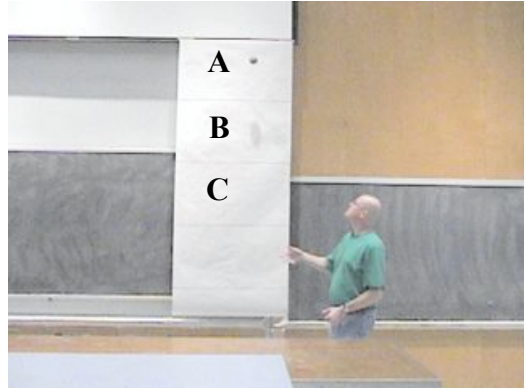

Assume the three regions where the ball is free from Roger's hand are labeled as in the thought experiment above (shown in the picture at right, but not on the actual video).

Play the video frame-by-frame, recording the number of frames the ball appears in each of the three regions:

| Region A | Region B | Region C |
|----------|----------|----------|
|          |          |          |

1. Determine the fraction and percentage of the frames that the ball is in each region. Use the table below to organize your results.

   |          | Fraction | Percentage |
   |----------|----------|------------|
   | Region A |          |            |
   | Region B |          |            |
   | Region C |          |            |

2. Compare your count and calculations with the other members of your group.

3. Is the sum of the percentages of time the ball is in Regions A, B, and C equal to what it should be? Resolve any discrepancies.

### VIII. Exploring an Oscillating Cart

We're going to do a similar experiment, but this one involves horizontal motion rather than vertical.

A.  Imagine a cart is placed on a flat surface, and connected to springs on either side. The region in which the cart travels in is divided into five equally spaced intervals, A–E. The equilibrium position of the cart is in the center of Region C. The arrow in the center of the cart is used as a marker to determine which region the cart is in. The cart is pulled back so that it is at the edge of Region A (shown in the diagram) and released. It oscillates back and forth.

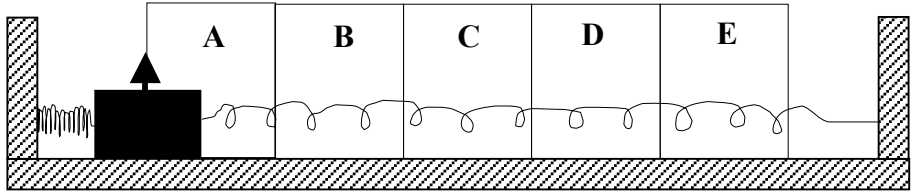

1. 100 photographs are taken at random times of the system. The photographs are then sorted by the region in which the marker on the cart appears. Which pile(s) of photographs, if any, will be the largest? Why?

2. You choose one of the photographs at random. In which region(s), if any, is it most likely to show the cart?

3. The experiment is restarted. After the cart is released, a single photograph is taken of the system at a random time. In which region(s), if any, is the cart most likely to be? Explain your reasoning.

4. Are 2 and 3 really the same question? Why or why not?

B.  Three students are discussing their reasoning about the piles of photographs.

   *Student 3: "Pile C will be the largest. Each time the cart oscillates back and forth, it passes twice through Region C."*

   *Student 4: "No, they'll all be about the same. Since the cart is traveling through equal-size regions at the same speed, it will spend the same amount of time in each region."*

   *Student 5: "Wait a minute – it is C, 'cause C is where it wants to be when it dies down."*

   1. <u>Answer this question individually</u>. With which, if any, of the statements do you agree? Explain your reasoning.

   2. Discuss the students' reasoning with the members of your group.

# Probability

C. Open the video clip *OscillatingCart.mov* on your computer. Play the clip, and observe an oscillating cart. For frames where the red dot appears on a line separating regions, decide on a method for assigning the position to one region or another.

1. How many frames show the red dot in each region? <u>Make sure you watch and count frames for the whole video!</u> Record your data in the chart below. (You'll have to establish criteria for deciding which region the dot is in when it appears on a boundary line.)

| Region A | Region B | Region C | Region D | Region E |
|---|---|---|---|---|
|  |  |  |  |  |

2. Plot the results in the space below.

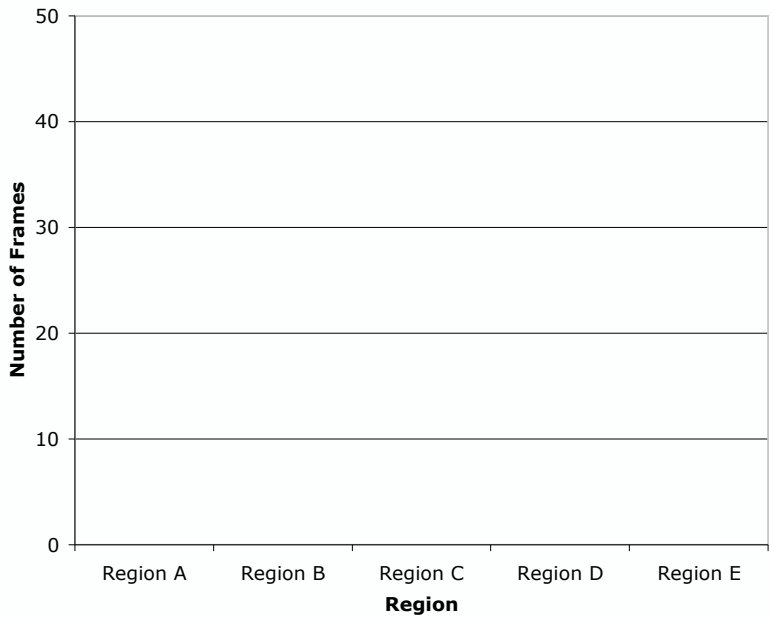

3. In which region(s) is the cart most likely to be found? Explain your reasoning.

4. In which region(s) is the cart least likely to be found? Explain your reasoning.

D. Re-examine the student dialogue in part B. As a group, re-evaluate their reasoning in light of your observation of the video. What could you tell each student to help them out?

   Student 3:

   Student 4:

   Student 5:

## IX. Board Meeting #2

Almost done! Take a minute to answer one of the following and share your ideas with your classmates.

1. This week you made histograms of coin flips, rolls of the dice, and the position of a cart over time. Compare and contrast these histograms to the histograms of electrons from last week's tutorial – that is, how are they alike? How are they different?

2. Imagine a scenario where a cart is rolling between two walls, as in the diagram below. There's no friction, so the car keeps rolling with constant speed, but when it hits a wall, it bounces back in the other direction going the same speed. Construct a histogram for this situation, and compare it to the histogram you developed for the oscillating cart.

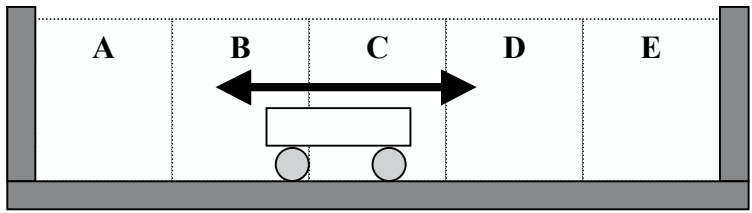

3. Revisit the thought experiment in VII, A, where 100 photographs were taken of the ball in flight. Imagine that we repeated that experiment 100 times, so there are now 100 piles with 100 photographs each (that's a lot of film!). At random, we select 1 photograph from each pile, making a new group of 100 photographs. How would the number of photographs showing the ball in Regions A, B, and C, respectively, in this pile compare with the number of photographs showing the ball in each region in a pile of 100 photographs from a single experiment?

Discuss the questions in class. Arrive at a consensus on each, and then go home!